\documentclass[twocolumn,showpacs,amsmath,amssymb]{revtex4}
\usepackage{graphicx}% Include figure files
\usepackage{dcolumn}% Align table columns on decimal point
\usepackage{bm}% bold math
\newcommand{\etal}{\emph{et al.}}
\newcommand{\be}{\begin{equation}}
\newcommand{\ee}{\end{equation}}
\newcommand{\bfig}{\begin{figure}}
\newcommand{\efig}{\end{figure}}
\newcommand{\incl}{\includegraphics}
\begin{document}      

\title{Divergent resistance at the Dirac point in graphene:
Evidence for a transition in a high magnetic field}

\author{Joseph G. Checkelsky, Lu Li$^*$ and N. P. Ong
}
\affiliation{
Department of Physics, Princeton University, Princeton, NJ 08544, USA
}
\date{\today}      
\begin{abstract}
We have investigated the behavior of the resistance of graphene
at the $n=0$ Landau Level in an intense magnetic field $H$.  
Employing a low-dissipation technique (with power $P<$3 fW), 
we find that, at low temperature $T$, the resistance at the
Dirac point $R_0(H)$ undergoes
a 1000-fold increase from $\sim$10 k$\Omega$ to 40 M$\Omega$ within 
a narrow interval of field.  The abruptness of the increase
suggests that a transition to an insulating, 
ordered state occurs at the critical field $H_c$.  
Results from 5 samples show that $H_c$ depends systematically
on the disorder, as measured by the offset
gate voltage $V_0$.  Samples with small $V_0$ display
a smaller critical field $H_c$.  Empirically, the steep increase in 
$R_0$ fits acccurately a Kosterlitz-Thouless-type correlation length 
over 3 decades.  The curves of $R_0$ vs. $T$ at fixed $H$
approach the thermal-activation form with a 
gap $\Delta\sim$15 K as $H\rightarrow H_c^{-}$, 
consistent with a field-induced insulating state.
\end{abstract}

\maketitle                   % Produces the title
\section{Introduction}\label{sec:intro}
In graphene, the low energy states display a linear
energy-momentum dispersion described 
by the Dirac Hamiltonian.  The observation of 
the integer quantum Hall (QH) effect by Novoselov \etal 
~\cite{Novoselov1,Novoselov2,Novoselov3} and 
Zhang \etal~\cite{Zhang1,Zhang2,Jiang} has sparked intense interest 
in this novel \emph{2D} (two-dimensional) system.
In a strong magnetic field $H$, the states are quantized into 
Landau Levels.  As a result of the Dirac dispersion,
the energy $E_n$ of the Landau Level (LL) (of index $n$)
increases with the flux density $B$ as $\sqrt{B}$, viz. 
$E_n = \mathrm{sgn}(n)\sqrt{2e\hbar v_F^2B|n|}$, where
$v_F$ is the Fermi velocity, $e$ the electron charge,
and $h$ is Planck's constant.  
The Hall conductivity is observed to be
accurately quantized as $\sigma_{xy} = (4e^2/h)[n+\frac12]=\nu e^2/h$,
where $e$ is the electron charge, $2\pi\hbar$ is Planck's constant,
and $\nu$ the sublevel index.  A key question 
is the nature of the ground state at the Dirac point.
In intense $H$, theory predicts interesting broken-symmetry
states driven by exchange and interaction.
These are characterized as quantum Hall Ferromagnetism (QHF)~\cite{MacDonald,Alicea,Yang,Abanin06,Goerbig,Ezawa},
or excitonic condensation~\cite{Khvesh,Miransky06,Gorbar,MiranskyEPL}.  
These collective states imply the existence of field-induced 
phase transitions, but the experimental situation is rather unsettled.  
Moreover, the proposed~\cite{Abanin06,Fertig,AbaninGeim} 
existence of counter-propagating 
edge modes at the Dirac point has further enriched the theoretical debate.
Is the high-field Dirac point a QH insulator or a QH metal?

Recently, we reported~\cite{Check} that the resistance 
at the Dirac point $R_0$ begins to increase steeply at $B$ = 10-12 T,
suggesting a transition to an insulating state. 
However, the results left open several key questions.
Because $R_0$ increased by only 1-decade (to 0.2 M$\Omega$)~\cite{Check},  
we could not establish that the high-field state is truly
insulating.  In graphene, the extreme 
sensitivity to thermal runaway has been highly problematical for 
researching its high-$H$ ground state~\cite{Zhang2,Jiang,Check}.
Adopting a low-dissipation technique to avoid self-heating,
we have measured the divergence in $R_0$ to 
40 M$\Omega$ ($\sim 1500\; h/e^2$) in 3 samples. 
Remarkably, the divergence 
is accurately described over 3 decades by the 
Kosterlitz-Thouless (KT) correlation length. 
The \emph{singular} nature of the divergence provides strong evidence
that a 2D transition to an insulating state occurs 
when $B$ exceeds a critical
field $H_c$.  The systematic variation of $H_c$ with $|V_0|$
(the gate voltage needed to bring the chemical potential $\mu$
to the Dirac point) implies that disorder is very effective in 
delaying $H_c$ to higher field values.  In all samples investigated
to date, the transition to the insulating state is reached
in fields below 35 T.

\section{Experimental Details}\label{sec:expt}
Empirically, problems associated with self-heating in graphene 
become serious when the power dissipated $P$ 
exceeds $\sim$10 pW for bath temperature
$T_b<$ 1 K.  As discussed in the Appendix, self-heating below 1 K
leads to a number of spurious features caused by thermal instability
in the sample.  We adopted a simple voltage-controlled technique
with ultra-low dissipation that avoids this difficulty, and allows
the divergence in $R_0$ to be measured reliably to 40 M$\Omega$.
An ac source maintains a fixed voltage amplitude (40 $\mu$V) 
across the sample in series with a 100-k$\Omega$ resistor
(details are given in the Appendix).
By phase-sensitive detection of both the current $I$ and 
the voltage $V_{xx}$, we have made 4-probe 
measurements of $R_0$ with ultra-low dissipation ($P$ 
decreases from $\sim$3 fW at 10 T to 40 aW above 25 T).  
Moreover, for $T<$1.5 K, the sample is immersed in 
liquid $^3$He so that the electrons in graphene are
in direct contact with the bath.  The largest reliably-measured $R_0$ 
is now 40 M$\Omega$ (limited by the input impedance 100 M$\Omega$
of the preamplifier).  
The new results provide an enlarged view of the 
interesting region in which $R_0$ diverges.  The samples
K52 and J24 have offset voltages $V_0$ much larger than that in K7,
the sample investigated in detail in Ref.~\cite{Check}.  
All samples except J18 were measured as-fabricated. Sample J18 
was subject to a $\frac12$ hr. anneal in He gas at 80 C
which decreased $V_0$ to 8 V.  However its large $H_c$ 
suggests that its initial value of $V_0$ (before annealing)
is very large.

In Samples K52, J18 and J24, the spacings $a$ between voltage leads are 
3.5, 2.75 and 3 $\mu$m, while the widths $w$ are
3, 3 and 2 $\mu$m, respectively.

%%%%%%%%%%%%%%%%%%%%%%%%%%%%%%%%%%%%% FIG
%%%%%%%%%%%%%%%%%%%%%%%%%%%%%%%%%%%%%%%
%%%%%%%%%%%%%%%%%%%%%%%%%%%%%%%%%%%%%%% 
\bfig[ht]			% Fig 1
\incl[width=7cm]{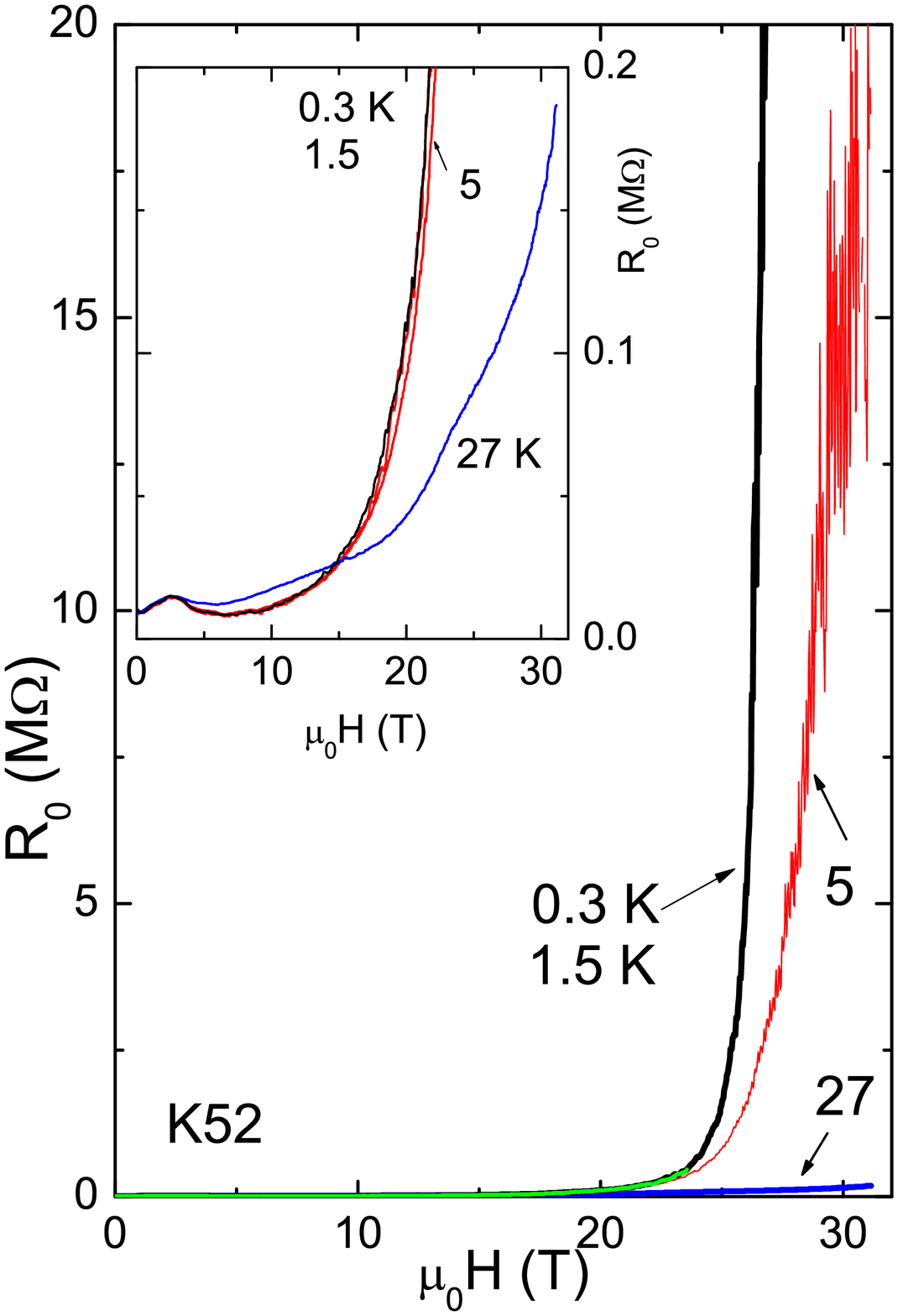}
\caption{\label{figRH} (color online)
(Main Panel) Divergence of the resistance $R_0$ at the Dirac point 
with $B$ at $T$ = 0.3, 1.5, 5 and 27 K (Sample K52).  
At 27 K, the increase in $R_0$ is quite moderate (to 190 k$\Omega$ 
at $H$ = 31 T).  At $T$ = 0.3 K, however, $R_0$ exceeds 
20 M$\Omega$ above 27 T.  
The curves at 0.3 and 1.5 K undergo a 1000-fold increase (40 k$\Omega$
to 40 M$\Omega$) in the narrow field interval 
17-27 T.  In high $B$, the 5 K curve deviates significantly 
from them. The inset shows the behavior of $R_0$ vs. $H$
in greatly expanded scale ($\times 100$).  
The voltage-regulated technique used for these measurements
dissipates $\sim$3 fW at 10 T and 40 aW above 25 T.
}
\efig

\section{Field dependence of $R_0$}\label{sec:field}
Figure \ref{figRH}  (main panel) shows curves of $R_0$ vs. $H$ in K52 at 
temperatures $T$ = 0.3 to 27 K.  As evident in the curves at 0.3 and 1.5 K,
$R_0$ undergoes a very steep, divergent increase when
$H$ exceeds $\sim$25 T.  The region just before the divergence occurs
is shown in greatly expanded scale in the inset.
At 27 K, the increase in $R_0$ is relatively modest
($\sim$20) between $B$ = 0 and 
31 T (inset).  However, as $T$ decreases to 5 K, the 
increase steepens sharply, as
reported~\cite{Check} for K7.  
Further cooling from 5 to 0.3 K changes the profile only very slightly.  
In the main panel, the curves at 0.3 and 1.5 K (which cannot be distinguished)
show that $R_0$ diverges to 40 M$\Omega$, with a slope that 
steepens rapidly with $H$.  The 3-decade increase (40 k$\Omega\rightarrow$
40 M$\Omega$) occurs within the narrow interval 17--27 T.
We find that the observed divergence is too steep 
to fit a power-law of the kind $R_0\sim (H_c-H)^{-\alpha}$,
with $\alpha>0$ and $H_c$ a critical field.

%%%%%%%%%%%%%%%%%%%%%%%%%%%%%%%%%%%%% FIG
%%%%%%%%%%%%%%%%%%%%%%%%%%%%%%%%%%%%%%%
%%%%%%%%%%%%%%%%%%%%%%%%%%%%%%%%%%%%%%% 
\bfig[ht]			% Fig 2
\incl[width=7.5cm]{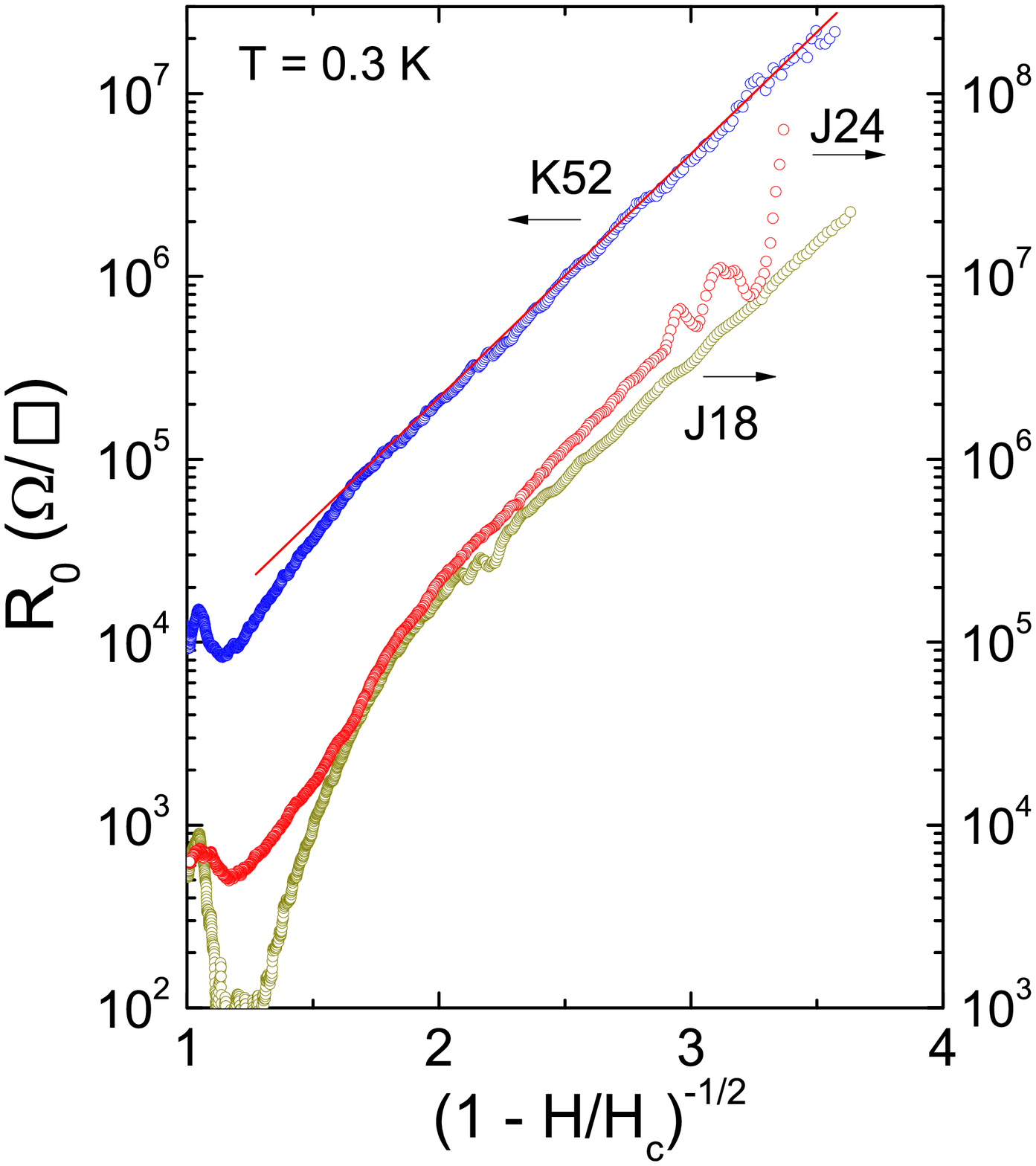}
\caption{\label{figKT} (color online)
Plot of $R_0$ vs. $1/\sqrt{1-h}$ 
for Samples K52, J18 and J24 at $T$ = 0.3 K, where
$h = H/H_c$ (the curve for K52 is displaced by 1 decade 
for clarity).  $R_0$ is expressed as
sheet resistance ($\Omega/\square$).  For each sample, 
$H_c$ is the optimal value that gives the best
straight-line fit vs. $1/\sqrt{1-h}$ (curvature is noticeable 
if $H_c$ is altered by $\pm$0.1 T from this value). 
For Samples K52, J18 and J24, $H_c$ equals 29.1, 
32.1 and 35.5 T, respectively. 
For K52, the thin solid line 
is the expression $R_{\xi}(h) = 440\exp[2b/\sqrt{1-h}]$, 
with $b$ = 1.54. The data match $R_{\xi}(h)$ very well
over nearly 3 decades in $R_0$.
}
\efig

As in Ref. \cite{Check}, we compare the divergence with that
predicted for the Kosterlitz-Thouless (KT) transition.
In \emph{2D} systems described by the XY model, the ordered phase is destroyed 
at the KT transition by unbinding of pairs of topological excitations
(e.g. vortices). As the transition here is induced by varying the applied field $H$,
we replace the reduced temperature $t$ by the reduced field $h=H/H_c$,
with $H_c$ the critical field.
In the limit $h\rightarrow 1^{-}$, the KT correlation length $\xi$ 
diverges as
\be 
\xi = a\exp[b/\sqrt{1-h}],
\label{KT}
\ee
where $a$ is the lattice parameter and $b$ a number $\sim 1$.

The quality of the fit to Eq. \ref{KT} is best
revealed in a plot of $\log R_0$ vs. the quantity
$1/\sqrt{1-h}$.  
In Fig. \ref{figKT}, we have plotted
$\log R_0$ in 3 samples K52, J19 and J24 against $1/\sqrt{1-h}$.
In each sample, the value of $H_c$ is adjusted to maximize
the high-field portion of the plot that falls on a straight line
(this is the only adjustment made).  
In Samples K52, J18 and J24, the inferred values of $H_c$ are 29.1, 
32.1 and 35.5 T, respectively. 
The values of $R_0$ in K52 fit the straight line 
representing the expression
$R_{\xi}(h) = 440\exp[2b/\sqrt{1-h}]$, 
with $H_c$ = 29.1 T and $b$ = 1.54.  The value of $b$ is 
consistent with the KT transition.  The 3-decade span is strong
evidence that Eq. \ref{KT} accurately describes the
divergence in $R_0$, and supports the inference that, at low $T$, 
we are observing a \emph{2D} KT-type phase transition to a high-field ordered state that is insulating.

%%%%%%%%%%%%%%%%%%%%%%%%%%%%%%%%%%%%% FIG
%%%%%%%%%%%%%%%%%%%%%%%%%%%%%%%%%%%%%%%
%%%%%%%%%%%%%%%%%%%%%%%%%%%%%%%%%%%%%%% 
\bfig[ht]			% Fig 3
\incl[width=8cm]{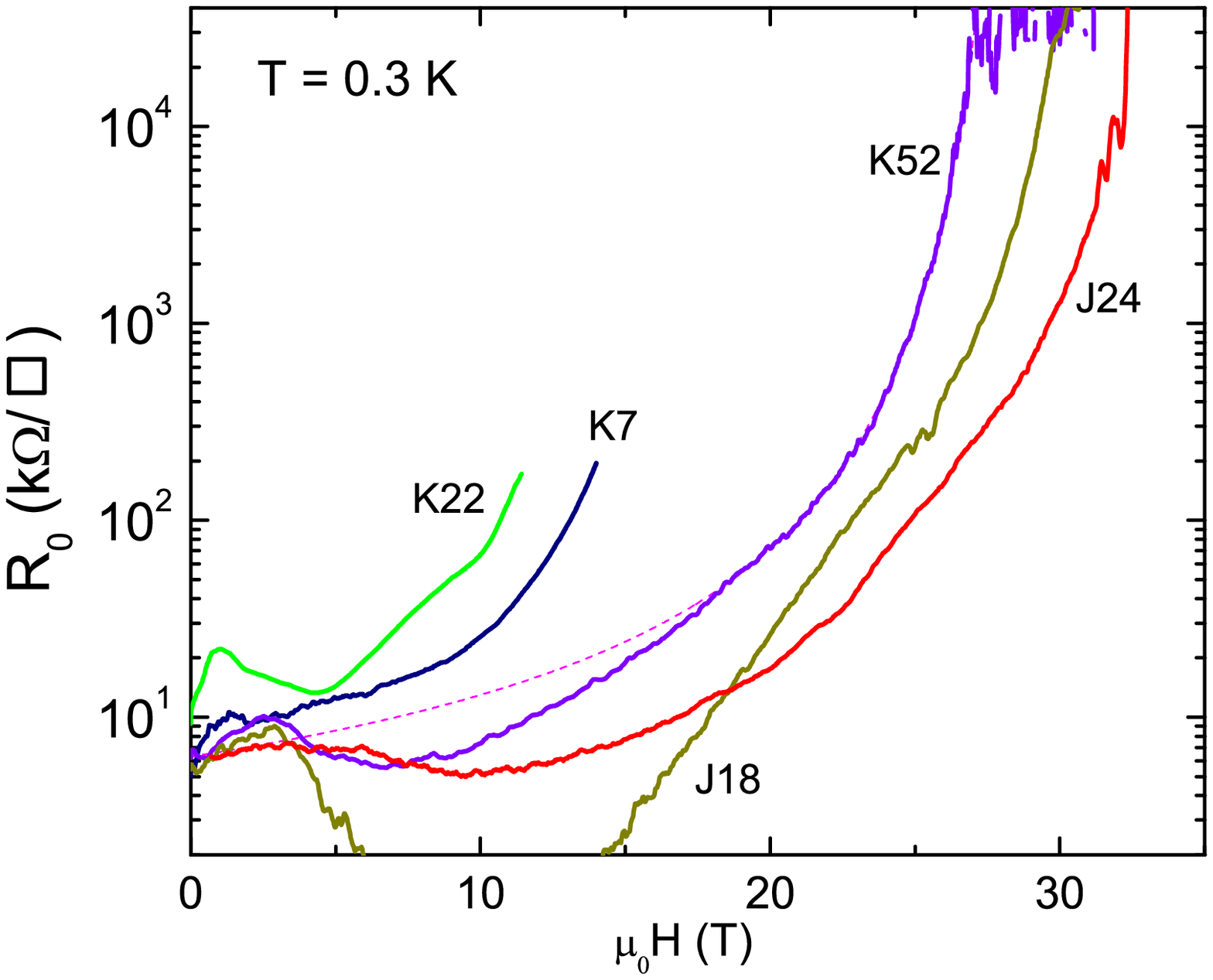}
\caption{\label{figRHcomp} (color online)
Comparison of the curves of $R_0$ vs. $H$ at 0.3 K
in log-log scale in 5 samples K7, K22, K52, J18 and J24
with gate voltage offsets $V_0$ = 1, -0.6, 3, 20, -- and 24 V, respectively
($V_0$ before annealing is not known in J18). 
$R_0$ is expressed as sheet resistance $\Omega/\square$.
In K7 and K22 (which have small $|V_0|$), the divergence 
in $R_0$ occurs at a lower $H$~\cite{Check}.  The
femtowatt-dissipation technique applied to K52, J18 
and J24 was not available for K7 and K22 (their curves
were limited to $R_0<$0.3 M$\Omega$).
The dashed curve is the fit of the K52 data to $R_{\xi}$.
}
\efig

The significant spread of $H_c$ inferred from the
fits in Fig. \ref{figKT} is in accord with Ref. \cite{Check}
which reported that $H_c$ correlates with the offset voltage
$V_0$.  In Fig. \ref{figKT}, the samples K52, J18 and J24
display a much larger 
critical field $H_c$ than the sample K7 (with $V_0$ = 1 V and
$H_c$ = 18 T) studied in detail
in Ref.~\cite{Check}.  Figure \ref{figRHcomp} plots together,
in log-log scale, the curves of $R_0$ vs. $H$ in the 
5 samples for which we have detailed high-field transport results.  
The systematic shift to higher fields of the divergence (in the order
K22, K7, K52 and J24) is matched by the increase
in their $V_0$ (-0.5, 1, 3, 20 and 24 V, respectively).  
The value of $V_0$ before annealing in J18 is not known.
The dependence of $H_c$ on $V_0$ is non-linear.  Initially
(for $0<V_0< 4$V), $H_c$ increases rapidly, but appears to
increase rather slowly when $V_0$ exceeds 20 V.

From a study of how $V_0$ affects the zero-field
transport, we have 
obtained evidence that the zero-field mobility $\mu_e$ 
is strongly suppressed if $V_0$ is large.    
In Fig. \ref{figV0}, we display curves of $R_{xx}$ vs. 
the unshifted gate voltage $V_g$ in a batch of 
samples that includes K22.  The curves are taken
at 295 K or 4 K (as indicated). In each sample, the 
width $\Delta V_g$ of the peak at the Dirac point 
directly measures $1/\mu_e$.
Clearly, the width increases dramatically with $V_0$.
These results support the inference that a large offset $V_0$ 
gives rise to large electronic disorder which enhances 
disorder scattering and suppresses $\mu_e$.
In turn, in an intense field, the transition field $H_c$ is 
pushed to higher values.  While these correlations do not exclude
other factors that may influence $H_c$, we have found that
$V_0$ is the single most reliable predictor of the 
field scale at which the divergence onsets at low $T$.

%%%%%%%%%%%%%%%%%%%%%%%%%%%%%%%%%%%%% FIG
%%%%%%%%%%%%%%%%%%%%%%%%%%%%%%%%%%%%%%%
%%%%%%%%%%%%%%%%%%%%%%%%%%%%%%%%%%%%%%% 
\bfig[ht]			% Fig 4
\incl[width=8cm]{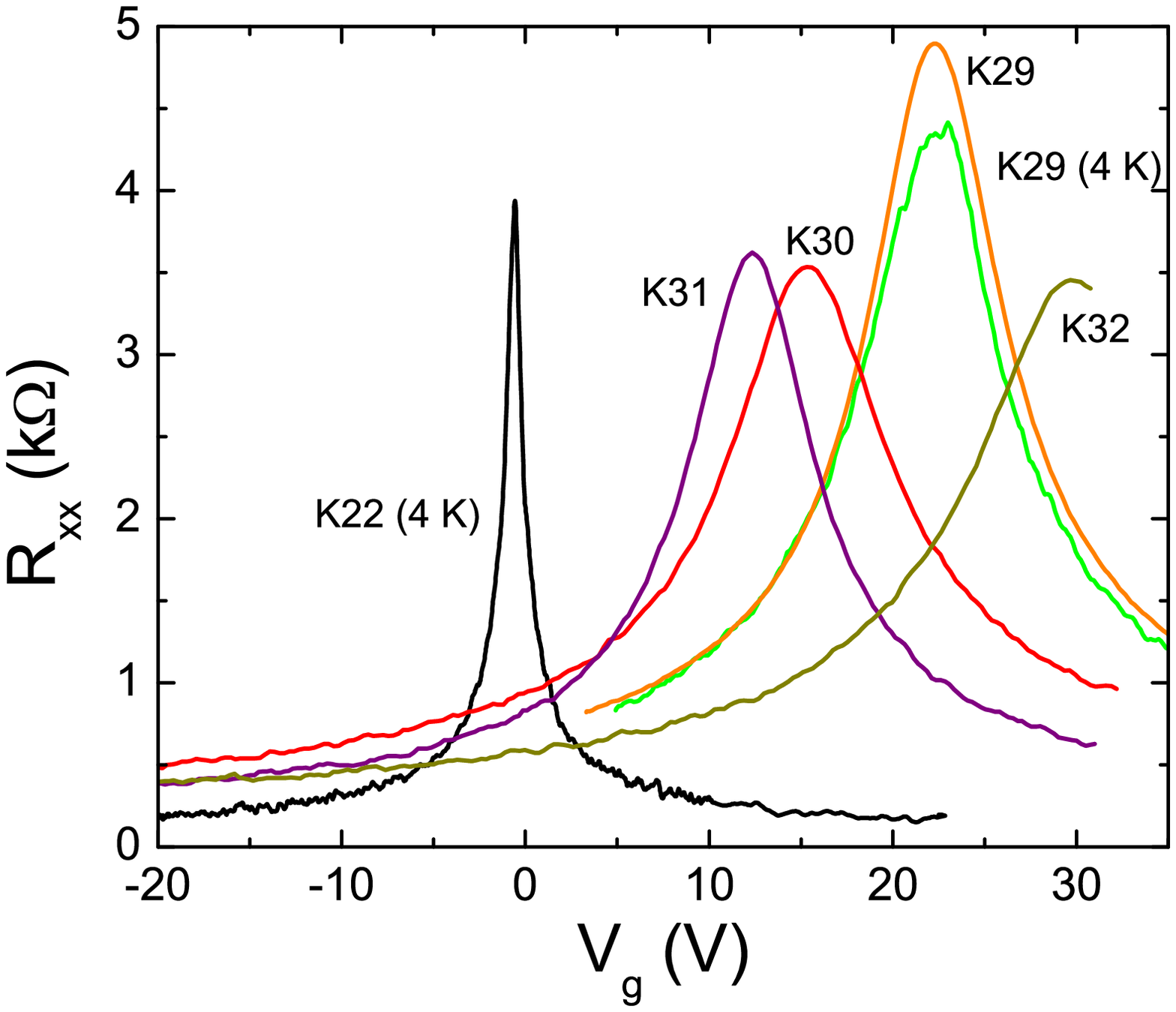}
\caption{\label{figV0}
Curves of the longitudinal resistance $R_{xx}$ vs. 
unshifted gate voltage $V_g$ in zero-$H$ in a series 
of samples at 295 K and 4 K (as indicated). The 
width of the peak in $R_{xx}$ increases systematically
with $V_0$ (located by the peak position).  This implies
that the mobility $\mu_e$ is very low in samples with large $V_0$.
}
\efig

%%%%%%%%%%%%%%%%%%%%%%%%%%%%%%%%%%%%% FIG
%%%%%%%%%%%%%%%%%%%%%%%%%%%%%%%%%%%%%%%
%%%%%%%%%%%%%%%%%%%%%%%%%%%%%%%%%%%%%%% 
\bfig[ht]			% Fig 5
\incl[width=8cm]{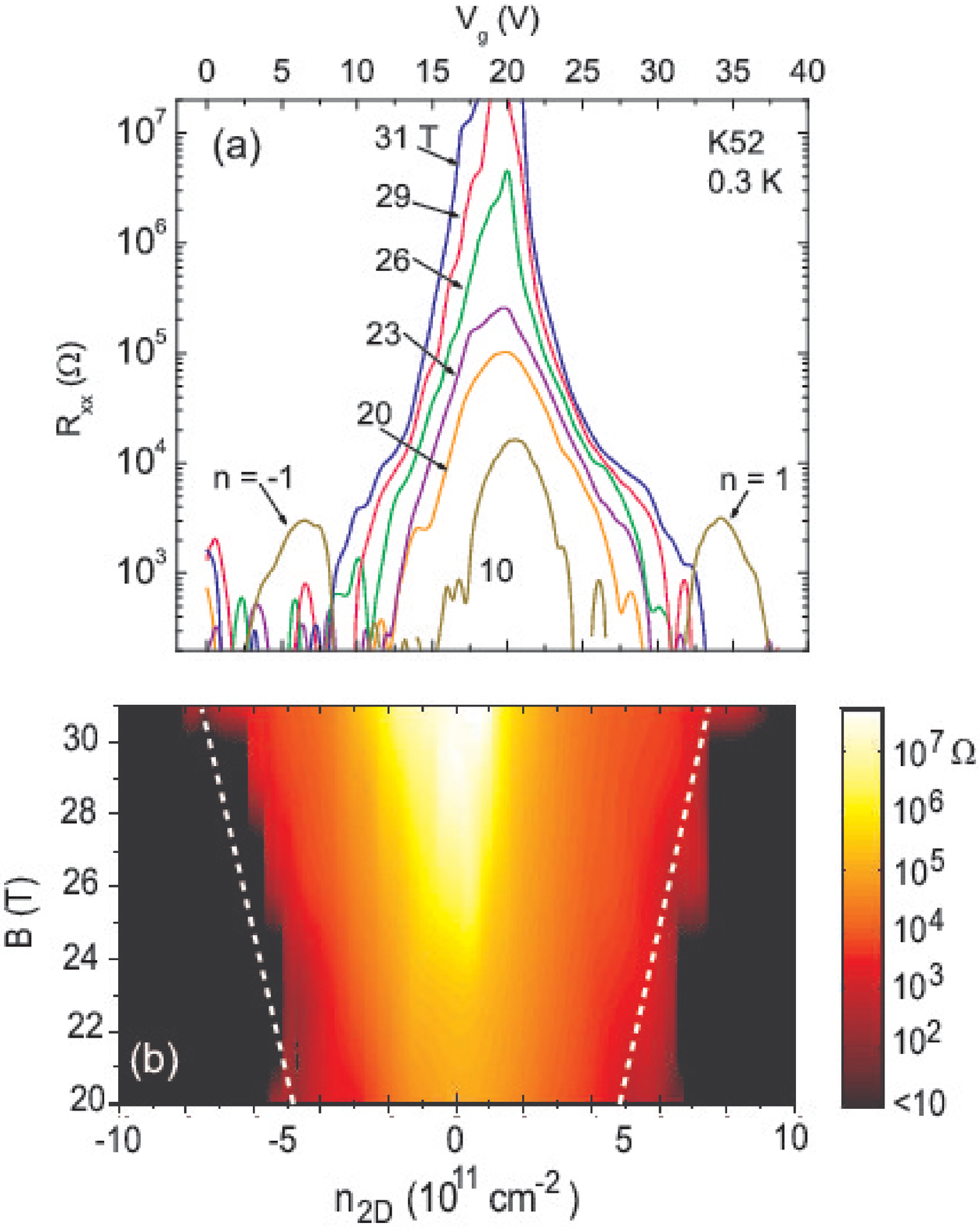}
\caption{\label{figRV} (color online)
Variation of $R_{xx}$ (Sample K52) vs. the gate 
voltage $V_g$ in the interval (0$< V_g < $ 38 V), with $B$ fixed at selected values 10--31 T, and $T$ = 0.3 K (Panel a). 
At $B$ = 10 T, the central peak ($n$ = 0 LL) is well 
separated from the LL peaks labelled as $n=\pm 1$.  
At $B$ = 20 T and higher, the $n=\pm 1$ LL's
fall outside the gate window.  With increasing $B$, 
the central peak increases rapidly and broadens.  
At the maximum $B$ (31 T), $R_{xx}$ at the Dirac point 
($V_g$ = 20 V) diverges to values above 40 M$\Omega$.  
The voltage-regulated technique has poor resolution when $R_{xx}$ falls 
below 0.3 k$\Omega$.  Panel (b) shows the contour plot of $R_{xx}(n_{2D},B)$
at 0.3 K in the $n_{2D}$--$B$ plane (color bar of $R_{xx}$ 
shown on right).  The dashed lines trace the sublevel degeneracy $1/2\pi\ell_B^2$. The density is given by 
$n_{2D} = CV_g/Ae$ where $C$ and $A$ are the capacitance and
area of the device, respectively ($C/A = 1.14\times 10^{-4}$ Fm$^{-2}$). 
}
\efig

%%%%%%%%%%%%%%%%%%%%%%%%%%%%%%%%%%%%% FIG
%%%%%%%%%%%%%%%%%%%%%%%%%%%%%%%%%%%%%%%
%%%%%%%%%%%%%%%%%%%%%%%%%%%%%%%%%%%%%%% 
\bfig[ht]			% Fig 6
\incl[width=7.5cm]{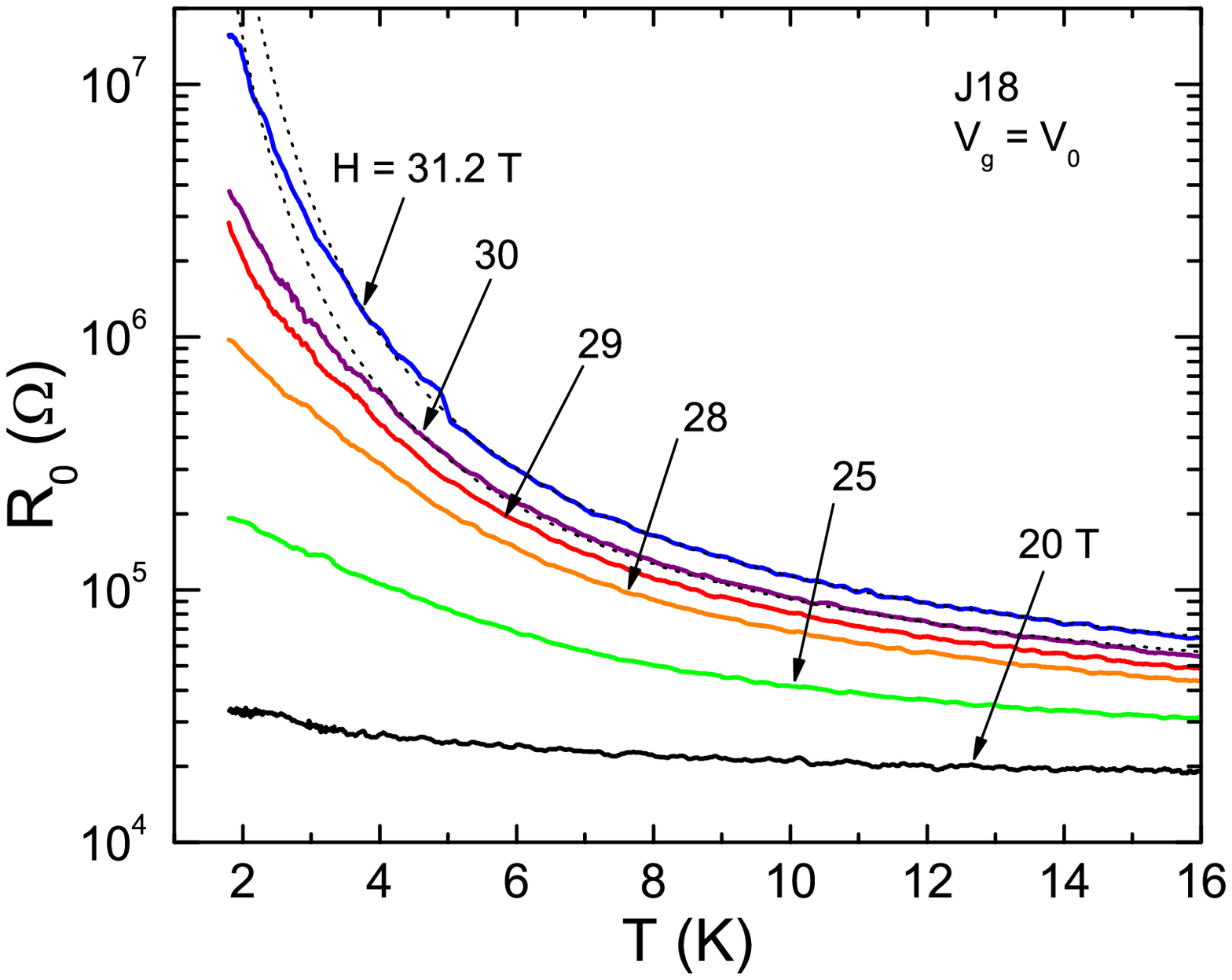}
\caption{\label{figRT} (color online)
The $T$ dependence of $R_0$ in Sample J18 with $H$
fixed at selected values in the interavl 20$<H<$ 31.2 T.
As $H$ approaches
$H_c$ (32.1 T), $R_0$ approaches the thermally activated
form ${\rm e}^{\Delta/T}$.  The thin dashed curves
are fits to this form with the gap $\Delta$ = 12.9 and 14.7 K
at $H$ = 30 and 31.2 T, respectively.
At the highest field (31.2 T), $R_0$ increases 
by a factor of 310 between 16 and 2 K.
}
\efig

\section{Doping dependence}
Further insight into the nature of the divergence is obtained by viewing
the behavior of the longitudinal resistance $R_{xx}$ vs. $V_g$ 
in a narrow gate window around the Dirac point at fixed $B$
(with $T$ kept at 0.3 K).  Figure \ref{figRV}a displays a series of 
curves of $R_{xx}$ (in log scale) vs. $V_g$ 
in K52 for fields $10\le H\le 31$ T.  
At 10 T, $R_{xx}$ displays 3 well-separated peaks corresponding to
$n=0$ LL at 20 V and the $n=\pm$1 LL's at 6 and 38 V, respectively.  
For $H\ge$ 20 T, the $n = \pm$1 levels move out of the gate-voltage 
window.  We focus on the divergent enhancement of the central peak 
as $B$ increases to 31 T.  The key feature is that $R_{xx}$ at 
the Dirac point ($V_g$ = 20 V) rises most
rapidly especially for $B>$25 T.

It is instructive to view $R_{xx}$ (at 0.3 K) as a contour plot in the
$n_{2D}$--$B$ plane where $n_{2D}$ is the \emph{2D} density 
of carriers doped by gating 
(Fig. \ref{figRV}b).  The color bar (right) 
gives the magnitude of $R_{xx}$.  Interestingly, the steep increase
in $R_{xx}$ with $B$ is confined to the region between the dashed lines,
which trace the sublevel degeneracy $1/2\pi\ell_B^2$.  This suggests that
only the states within the lowest sublevels (on either side of $\nu = 0$)
are affected by the opening of the gap $\Delta$.  
The contours appear to converge to a rounded
cusp at $n_{2D} =0$, but with a curvature that increases rapidly with $B$.  
At the largest $R_{xx}$ (white region), the contour resembles a
a narrow, sharp wedge.  The contour pattern suggests 
that $H_c$ increases very rapidly
from the value 29.1 T, as $|n_{2D}|$ deviates from 0.

The physical picture implied by the results is that, at the Dirac point, 
a field-induced transition to a gapped, insulating state occurs at $H_c$.  
The value of $H_c$ is highly sensitive to slight deviations away from 
exact charge neutrality (Fig. \ref{figRV}).  As we decrease
$H$ below $H_c$, the ordered state is unstable to the spontaneous unbinding
of (vortex-like) topological excitations which have a
mean spacing of $\xi$ (Eq. \ref{KT}). 
Because $R_0$ fits accurately to $\xi^2$ over
3 decades, we infer that the conductance scales as the
density of excitations.  Hence the excitations are charged,
and they carry the entire current $I$ in the 
limit $T\rightarrow 0$.  As this conduction 
channel is qualitatively distinct from thermally activated carriers,
it may account for the unusual ``saturation'' behavior of $R_0$ vs. $T$~\cite{Check}.
As $T\rightarrow 0$, with $H$ fixed near $H_c$, the conduction crosses 
over at 1 K from a steep, thermally
activated channel to a $T$-independent channel carried by the
excitations (Fig. \ref{figRH}b).

\section{Temperature dependence}\label{sec:temp}
The $T$ dependence of $R_0$ in J18 with $H$ fixed at selected values
is displayed in semilog scale in Fig. \ref{figRT}.
At the relatively high field $H$ = 20 T, the 2-fold increase
in $R_0$ between 16 and 2 K is quite modest. However, above 20 T, 
the $T$ dependence steepens rapidly.  As $H$ is increased towards
the critical $H_c$ (=32.1 T), the profile of $R_0$ moves ever closer to the 
activated form $R_0\sim \mathrm{e}^{\Delta/T}$.  
We have plotted fits to the activated
form (dashed curves) at the 2 highest fields $H$ = 30 and 31.2 T.
At the highest $H$, the measured $R_0$ tracks closely the dashed curve until
$T$ falls below 3 K where it deviates downwards.  As $H$ decreases
further from $H_c$, the deviations start at a higher $T$.
The gap value $\Delta$ equals 12.9 and 14.7 K at 30 and 31.2 T, respectively.  
The latter provides an estimate of the energy
gap in the ordered state for J18.  Establishing that $R_0$ is
thermally activated when $H>H_c$ is an essential goal, as it shows that 
the ground state above $H_c$ is a true insulator with a well-defined
gap order parameter (as opposed to a state in which the carriers
are strongly localized).  Although the curves in Fig. \ref{figRT} come close to
establishing this result, measurements of $R_0$ vs. $T$ in samples with 
more accessible $H_c$ are desirable.

Interestingly, throughout the 
pre-transition region ($20<H<H_c$), $R_0$ also displays large increases
with decreasing $T$.  The carriers are strongly affected by the
impending insulating state.  We interpret these changes as reflecting very 
strong fluctuations in the order parameter that characterizes
the ordered insulating phase.  The strong $T$ dependence is also 
apparent in the fixed-$T$ curves shown in Fig. \ref{figRH}.

Quite apart from the thermal activation argument, there are other
evidence to suggest that the divergence is not consistent
with electron localization.  As evident from measuring the 
widths of the peaks of $R_{xx}$ vs. $V_g$ taken in zero $B$,
samples with smallest $|V_0|$ are the least disordered.  
The electron mobility $\mu_e$ decreases from $\sim$2.5  
to 0.5 T$^{-1}$ as $|V_0|$ increases from 0.5 to 20 V.
With this trend in mind, we compare in Fig. \ref{figRHcomp} 
the profiles of $R_0$ vs. $B$ 
in K52 with K7 and K22~\cite{Check}.
In Samples K22 and K7 (with $V_0$ = -0.6 and 1 V, respectively), the
divergence in $R_0$ is apparent in relatively low $B$ (below 12 T).
By contrast, we must go to much higher fields ($>$20 T) in K52.  
The dashed line is the fit to $R_{\xi}$ in K52 ($H>$ 18 T) described above.

In the localization scenario, the observed divergence of $R_0$ in strong field
is explained by postulating that $B$ induces localization of 
the electrons.  However, applying this reasoning 
to the 3 samples in Fig. \ref{figRHcomp}, we would conclude that a modest $B$
is sufficient to trigger the localization in clean samples,
but very intense fields are needed in dirtier samples.  
This implies that disorder and field act in opposition to bring
about localization, which is in conflict with physical intuition.
In addition, localization induced by $B$ cannot lead to the singular 
divergence observed in $R_0$ (Fig. \ref{figKT}).
For these reasons, we believe that localization is not a viable
explanation for the divergence in $R_0$.  

\section{Discussion}\label{sec:discuss}
In the absence of Zeeman splitting and electron-electron interaction, the $n=0$ LL is 4-fold degenerate corresponding to the
(physical) spin degeneracy indexed by $\sigma=\pm 1$ and the
$K$ and $K'$ valley degeneracy indexed by $\tau=\pm 1$.  
At $\nu=0$, the energy of the 
$n=0$ LL, $E_s^{\tau}$, is zero.  The effect of 
interaction in producing a broken-symmetry ground state has been
investigated by several groups.  To discuss our experiment,
it is convenient to distinguish 2 different theoretical 
scenarios for the $n=0$ LL.

In one scenario, the Quantum Hall Ferromagnet (QHF) models, the 
exchange energy $E_{ex}\sim\sqrt{B}$ 
leads to ferromagnetic polarization of the physical spins
~\cite{MacDonald,Goerbig,Alicea,Abanin06,Fertig}.  This
produces a spin gap in the bulk without affecting the valley degeneracy, i.e.
$E_\sigma^{\tau} = \sigma(\mu_B B + E_{ex})$ with $\mu_B$ the Bohr magneton.  Near the edge of the sample, the residual valley degeneracy is lifted by the edge potential.
An important consequence of the QHF scenario at $\nu=0$ is the existence of 
spin-filtered counter-propagating edge (CPE) modes 
which result in a residual conductance of $2e^2/h$ regardless
of the magnitude of the spin gap in the bulk ~\cite{Abanin06,AbaninGeim,Fertig}.
[In principle, the CPE modes are not present if the exchange 
polarizes instead the valleys to produce the bulk gap
(this involves the same exchange energy $E_{ex}$).  
However, most investigators favor the spin-polarization scenario in
graphene because it is augmented
by the Zeeman energy $\mu_BB$ (the valley-polarization scenario is 
also called the QHF following the original usage~\cite{Moon} in quadratic,
bilayer GaAs-based devices).]  

In the second scenario, called magnetic catalysis~\cite{Khvesh,Miransky06,Gorbar,MiranskyEPL,Ezawa}, the field component $B_{\perp}$ normal to the graphene sheet triggers
electron-hole condensation.  The instability introduces a mass term
to the Dirac equation which leads to the order parameter~\cite{Khvesh,Miransky06,Gorbar}
$\langle\bar{\Psi}\Psi\rangle = \sum_{\tau\sigma} (|\psi_{\tau A\sigma}|^2 -
|\psi_{\tau B\sigma}|^2)$, 
where $\psi_{\tau A\sigma}$ and $\psi_{\tau B\sigma}$ are the
wave functions of electrons of spin $\sigma$ and valley $\tau$ 
at sites A and B, respectively.  The instability -- a solid-state realization of chiral-symmetry breaking in \emph{(2+1)D}~\cite{Miransky94} -- results
in preferential occupation of, say, the A sublattice sites over the
B sites, and drives the system into an insulating state.  
Significantly, the instability is strongest for $n=0$.

The steep increase in $R_0$ vs. $H$ first reported in Ref. \cite{Check} implies
that at large $H$, the ground state at the Dirac point 
has a resistance at least 20 times larger than the quantum $h/e^2$.  
Although the measurements were limited to $R_0<$ 0.3 M$\Omega$, 
the upturn appeared to diverge at a critical field $H_c$, suggestive of
a singular field dependence.  The findings are clearly at odds with the existence
of CPE modes (see, however, the results in Ref. \cite{AbaninGeim}).

In the present report, we have extended by a factor of 200 the range 
of resistance measurements and shown that, at 0.3 K, the increase in $R_0$ is truly
divergent as well as singular.  Moreover, this behavior has been observed in
all samples investigated to date by us in high fields.  The evidence amassed clearly establish that the high-field ground state 
at the Dirac point is a true insulator
(at least for samples prepared on a SiO$_2$ substrate).    
Ipso facto, the CPE modes do not exist in the insulating state.
However, our results do not preclude them at low fields.

Lately, several groups have considered how the CPE modes are affected by
intense field. A very interesting possibility is that an intense field
destroys the CPE modes in a field-induced transition.  
It has been pointed out to us that the CPE modes are not 
protected against 2-particle exchange scattering with 
spin flip~\cite{Haldane}.  As the exchange energy increases with $B$,
the increased scattering rate could lead to a gap in the edge modes.  

In the magnetic catalysis scenario, Gorbar \etal~\cite{Gorbar}
recently considered
the competition between the mass gap $\langle\bar{\Psi}\Psi\rangle$ and the
spin gap (augmented by Zeeman energy) and inferred that CPE
states exist only above a critical field $B_{cr}$.

Shimshoni \etal~\cite{Shimshoni} have proposed that scattering off 
magnetic impurities can lead to strong 
localization of electrons in CPE modes which can mimic a KT transition.  
However, this scenario needs to be reconciled with the observed nearly
activated behavior of $R_0$ as well as the variation of
$H_c$ with $V_0$ in different samples.

\section{Appendix}
Near the Dirac point, resistance traces are strongly distorted
when the Ohmic heating $P$ exceeds 10 pW at bath temperatures $T_b$ below 1 K.
As examples, we plot in Fig. \ref{figheat} resistance traces (with $I$ fixed).
In Panel a, the inferred curve of $R_{xx}$ vs. $V_g$,
measured with $B$ = 31 T, $I$ = 10 nA and $T$ = 0.3 K, shows a pronounced dip 
near the Dirac point caused by self-heating instability (the true 
$R_0$ exceeds 10 M$\Omega$).  Panel b shows ``$R_0$'' vs $B$ measured at fixed $I$.
In the curve for K23 (at 5 K), self-heating reverses the
trend of $R_0$.  The downturn is avoided when 
$I$ is decreased to 1 nA, but the measured curve (in K22 at
0.3 K) is still greatly suppressed from the true divergent profile.

%%%%%%%%%%%%%%%%%%%%%%%%%%%%%%%%%%%%% FIG
%%%%%%%%%%%%%%%%%%%%%%%%%%%%%%%%%%%%%%%
%%%%%%%%%%%%%%%%%%%%%%%%%%%%%%%%%%%%%%% 
\bfig[h]			% Fig A1
\incl[width=7cm]{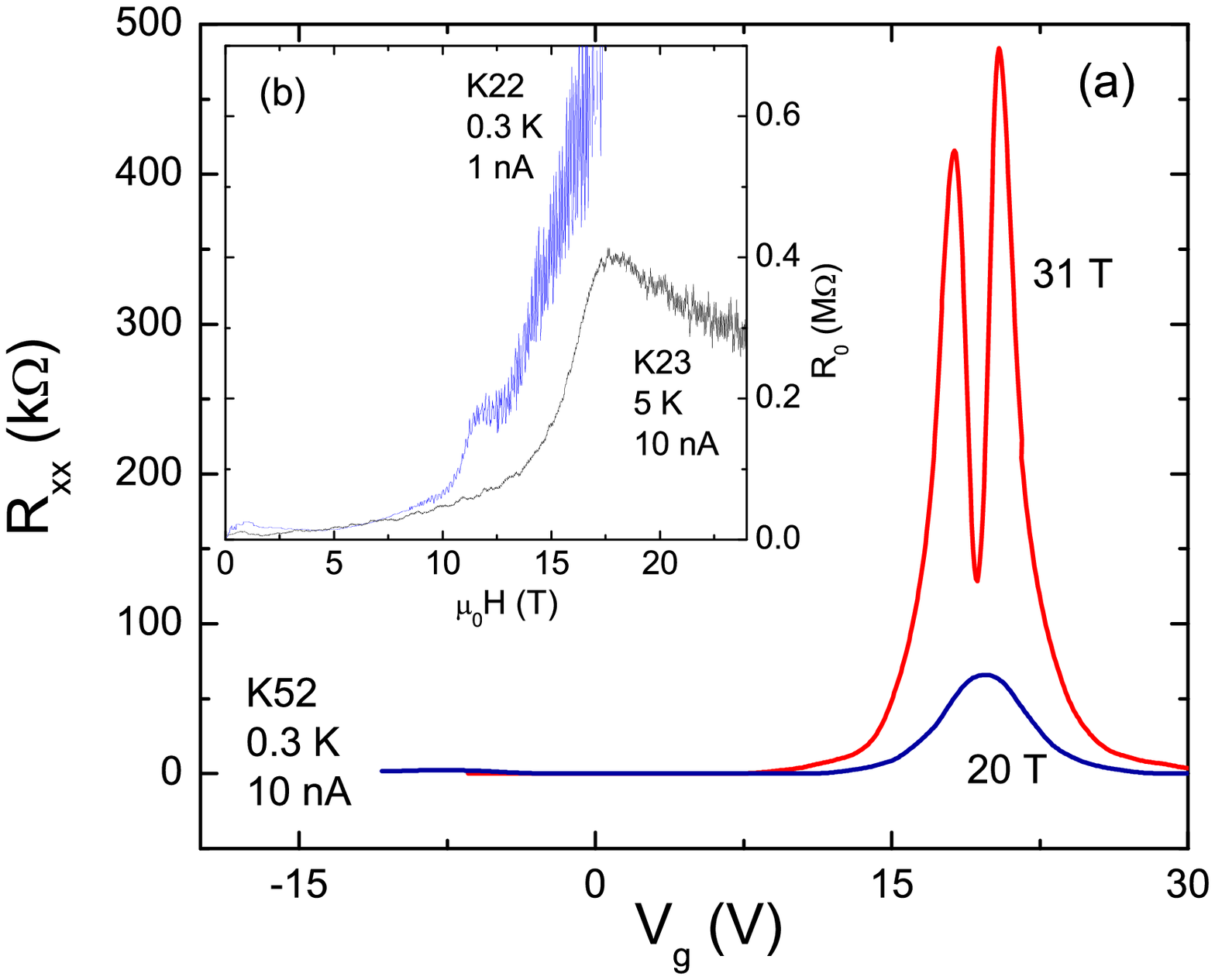}
\caption{\label{figheat} (color online)
Spurious features caused by sample self-heating in graphene.
Panel (a) shows a gate-sweep measurement at $T$ = 0.3 K
of $R_{xx}$ vs. $V_g$ in K52,
with $I$ fixed at 10 nA (dc).  At $B$ = 31 T
(red curve), severe self-heating inverts
the peak at the Dirac point (as shown in Fig. \ref{figRH},
$R_0$ actually exceeds 40 M$\Omega$).
Heating effects are less severe in the 20-T curve.
Panel (b) displays curves of ``$R_0$'' vs. $B$ measured in Samples 
K22 (at 0.3 K) and K23 (at 5 K) with $I$ fixed at 1 and 10 nA (dc), respectively.
When $R_0>$ 0.2 M$\Omega$, self-heating produces 
the spurious shoulders and broad peaks,
whose positions and shapes depend on $I$.
}
\efig

%%%%%%%%%%%%%%%%%%%%%%%%%%%%%%%%%%%%% FIG
%%%%%%%%%%%%%%%%%%%%%%%%%%%%%%%%%%%%%%%
%%%%%%%%%%%%%%%%%%%%%%%%%%%%%%%%%%%%%%% 
\bfig[ht]			% Fig A2
\incl[width=7cm]{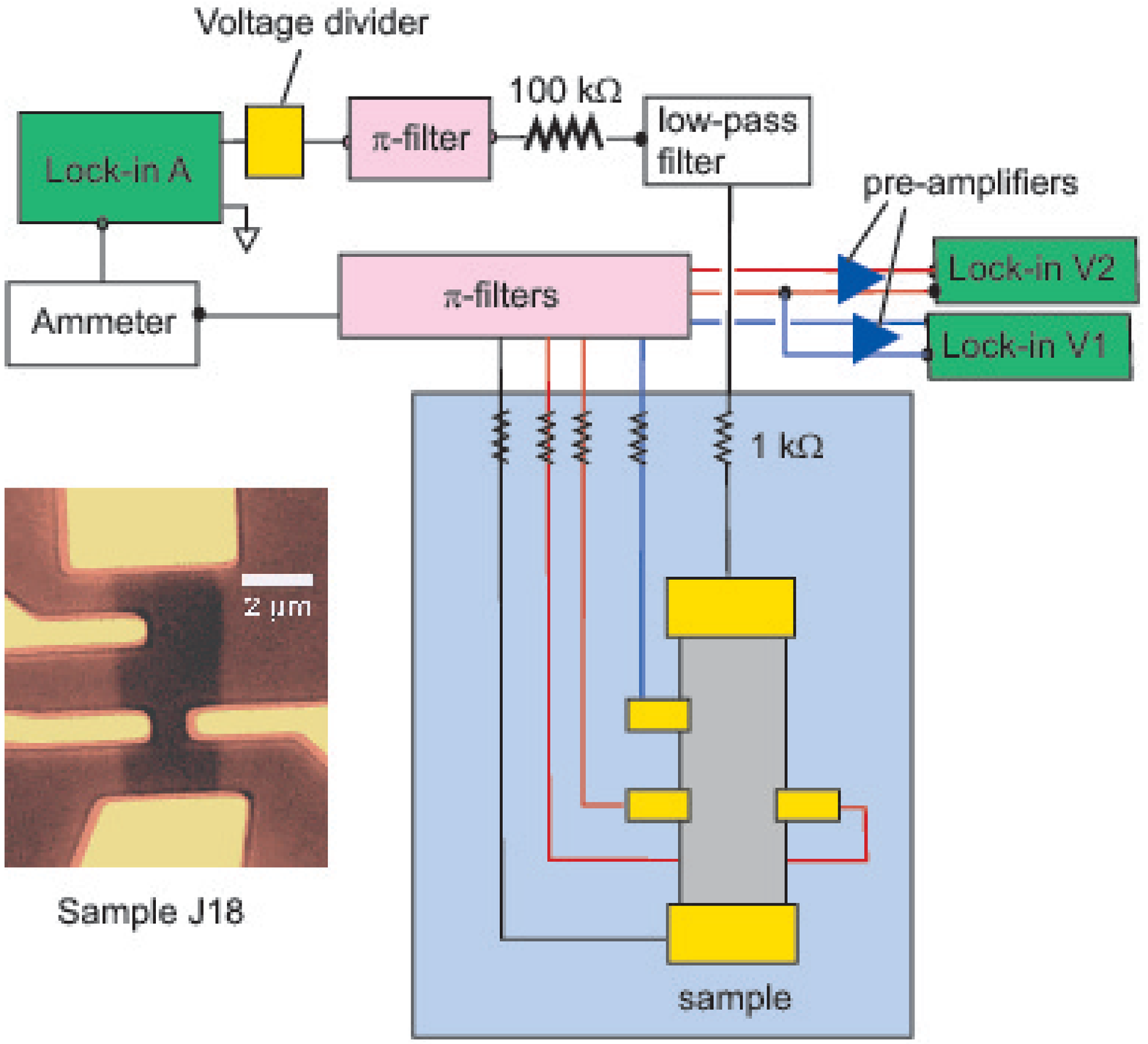}
\caption{\label{figcircuit} (color online)
Schematic of the low-dissipation, voltage-regulated circuit used in the
experiment.  Lock-in (amplifier) A produces a regulated
voltage emf (3 Hz) that is reduced to an amplitude of 40 $\mu$V
by a 100:1 voltage divider.  The signal goes through a $\pi$ filter,
a buffer resistor (100 k$\Omega$) and a low-pass filter 
before entering the dewar.
The AC current passing through the graphene sample is measured
by the picoammeter (Keithley), whose output is phase-detected by
Lock-in A. The longitudinal voltage $V_{xx}$ and Hall voltage 
$V_{xy}$ are phase-detected by Lock-ins $V_1$
and $V_2$, respectively, after transmission through 
a bank of $\pi$-filters and high-impedance (100 M$\Omega$)
pre-amplifiers.  As shown, all wires
entering the dewar are buffered by 1 k$\Omega$ nichrome
thin-film ceramic resistors.  The inset (lower left) shows
Sample J18.
}
\efig

Figure \ref{figcircuit} is a schematic of the measuring circuit 
employed in the ultralow-dissipation technique.  A nominally constant ac voltage
($\sim$40 $\mu$V) of frequency 3 Hz is applied across the sample in series with a 100-k$\Omega$ buffer resistor.  The current passing through the sample
is measured by a Keithley picoammeter whose output is phase-detected by the
lock-in amplifier A.  Simultaneously, the longitudinal voltage $V_{xx}$
and Hall voltage $V_{xy}$ are phase-detected by 2 other lock-ins.  As shown,
all wires entering the dewar are filtered and buffered to exclude extraneous
RF signals which may be a potential source of sample heating.

%%%%%%%%%%%%%%%%%%%%%%%%%%%%%%%%%%%%% FIG
%%%%%%%%%%%%%%%%%%%%%%%%%%%%%%%%%%%%%%%
%%%%%%%%%%%%%%%%%%%%%%%%%%%%%%%%%%%%%%% 
\bfig[h]			% Fig A3
\incl[width=8cm]{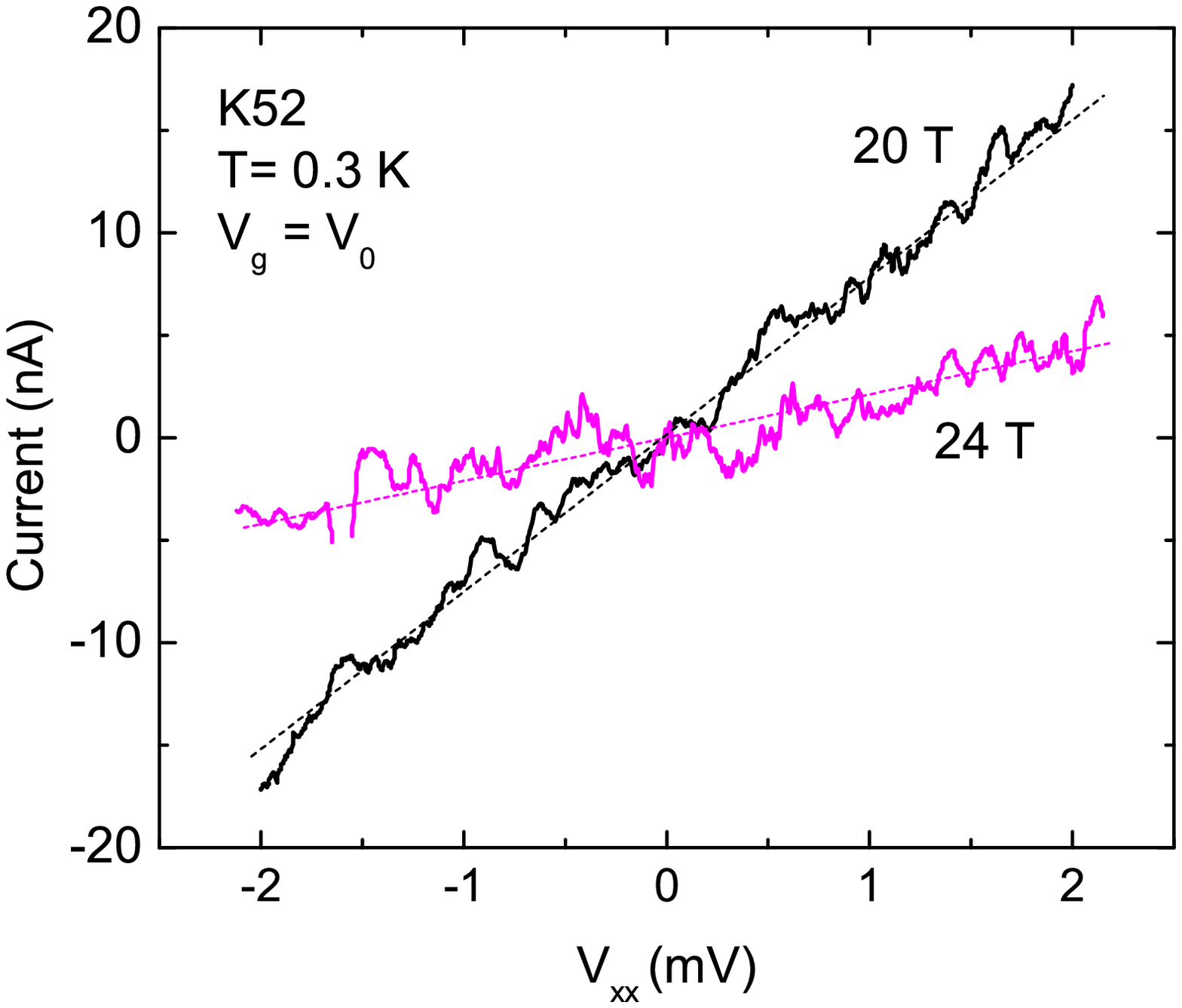}
\caption{\label{figIV} (color online)
The current-voltage curves measured at the Dirac Point ($V_g=V_0$)
in Sample K52 at $T$ = 0.3 K at 2 the fields $H$ = 20 and 24 T
(using the circuit in Fig. \ref{figcircuit}).
The linearity of $I$ vs. $V_{xx}$
implies that, at these fields, self-heating is not 
observable at these power-dissipation levels
(the dashed lines are guides to the eye).
}
\efig

Using the ultralow-dissipation technique, we completely avoid
the thermal runaway problems illustrated in Fig. \ref{figheat}.
At selected fields, we have performed $I$-$V$ measurements
to check that self-heating is not skewing the results even at our 
lowest $T$ (0.3 K).  Figure \ref{figIV}
shows curves of $I$ vs. $V_{xx}$ at the Dirac point in K52 at $T$ = 0.3 K
with $H$ fixed at 20 and 24 T.  The linearity implies that self-heating
is not observable up to a bias voltage of 2 mV.  Since all the 
curves displayed in the main text were taken with a bias of 40 $\mu$V, we are
comfortably within the Ohmic regime.

%%%%%%%%%
%%%%%%%%%%
We thank H. A. Fertig, P. A. Lee, D. Abanin, V. A. Miransky, 
E. Shimshoni, F. D. M. Haldane and D. N. Sheng for
valuable comments, and acknowledge support from NSF-MRSEC under Grant DMR-0819860,
and from the Princeton Center for Complex Materials.  The experiments were performed at the National High Magnetic Field Laboratory, which is supported by NSF Cooperative Agreement No. DMR-084173, by the State of Florida, and by the Department of Energy. 
%%
%%%%%%%%%%%%%%%%%%%%%%%%%%%%%%%%%%%%%% 
%%%%%%%%%%%%%%%%%%%%%%%%%%%%%%%%%%%%%%% 
\vspace{0.2cm}\\
$^*$\emph{Current address of LL}: Dept. of Physics, MIT

\end{document}